\documentstyle[12pt]{article}
\makeatletter
\renewcommand\thesection{\@Roman\c@section}
\renewcommand\thesubsection{\thesection.\@arabic\c@subsection}
\makeatother

\newcommand{\sect}[1]{\setcounter{equation}{0}\section{#1}}

\begin{document}



\def\R{\overline{R}}

\newtheorem{Theorem}{Theorem}
\newtheorem{Definition}{Definition}
\newtheorem{Proposition}{Proposition}
\newtheorem{Lemma}{Lemma}
\newtheorem{Corollary}{Corollary}
\newcommand{\proof}[1]{{\bf Proof. }
        #1\begin{flushright}$\Box$\end{flushright}}

\baselineskip=20pt

\newfont{\elevenmib}{cmmib10 scaled\magstep1}
\newcommand{\preprint}{}
\newcommand{\Title}[1]{{\baselineskip=26pt
   \begin{center} \Large \bf #1 \\ \ \\ \end{center}}}
\newcommand{\Author}{\begin{center}
   \large \bf
Shao-You Zhao${}^{a,b}$, Wen-Li Yang${}^{a,c}$ and ~Yao-Zhong
Zhang${}^a$
 \end{center}}
\newcommand{\Address}{\begin{center}

     ${}^a$ Department of Mathematics, University of Queensland,
            Brisbane, QLD 4072, Australia\\
     ${}^b$ Department of Physics, Beijing Institute of
     Technology, Beijing 100081, China\\
     ${}^c$ Institute of Modern Physics, Northwest University,
     Xian 710069, P.R. China\\\vskip.1in

 E-mail: syz@maths.uq.edu.au, wenli@maths.uq.edu.au, yzz@maths.uq.edu.au

 \end{center}}
   \newcommand{\Accepted}[1]{\begin{center}
   {\large \sf #1}\\ \vspace{1mm}{\small \sf Accepted for Publication}
   \end{center}}
\preprint \thispagestyle{empty}
\bigskip\bigskip\bigskip

\Title{Determinant Representation of Correlation Functions for the
$U_q(gl(1|1))$ Free Fermion Model}  \Author

\Address \vspace{1cm}

\begin{abstract}

With the help of the factorizing $F$-matrix, the scalar products
of the $U_q(gl(1|1))$ free fermion model are represented by
determinants. By means of these results, we obtain the determinant
representations of correlation functions of the model.

\vspace{1truecm} \noindent {\it PACS:}~~ 04.20.Jb, 05.30.-d,
71.10.Fd

\vspace{1truecm} \noindent {\it Keywords}:  Correlation function,
Drinfeld twist, Algebraic Bethe ansatz.
\end{abstract}

\newpage
\sect{Introduction}

The computation of correlation functions is one of the challenging
problem in the theory of quantum integrable lattice models
\cite{Korepin82,Korepin93}. In this paper, we compute the
correlation functions of the free fermion model by means of the
algebraic Bethe ansatz method
\cite{Korepin82,Korepin93,Izergin87}. Our computation are based on
the recent progress on the Drinfeld twists. Working in the
$F$-bases provided by the $F$-matrices (Drinfeld twists), the
authors in \cite{Maillet96,Kitanine98} managed to derive the
determinant representations of the form factors and correlation
functions of the XXX and XXZ models in the framework of algebraic
Bethe ansatz.

Recently we have constructed the Drinfeld twists for both the
rational $gl(m|n)$ and the quantum $U_q(gl(m|n))$ supersymmetric
models and resolved the hierarchy of their nested Bethe vectors in
the $F$-basis \cite{zsy04,zsy0502,zsy0503}. These results serve as
the basis of our computation in this paper of the correlation
functions of the $U_q(gl(1|1))$ free fermion model.

Correlation functions of the free fermion model based on the XX0
spin chain (XY model \cite{Lieb61}) with periodic boundary
condition were studied in \cite{McCoy68}-\cite{Kitanine02}. As is
seen in section VI, by using the Jordan-Wigner transform, our
$U_q(gl(1|1))$ free fermion model is equivalent to a twisted XX0
model, and the one-point functions we obtained (see (\ref{eq:F-})
and (\ref{eq:F+}) below) give the $m$-point correlation functions
of the twisted XX0 model (see e.g. (\ref{eq:F-XX0})).

The present paper is organized as follows. In section II, we
review the background of the $U_q(gl(1|1))$ model and its
algebraic Bethe ansatz. In section III, we construct the Drinfeld
twists for the model. In section IV, we obtain the determinant
representation of the scalar products of the $U_q(gl(1|1))$ Bethe
states. Then in section V, we compute correlation functions of the
local fermionic operators of the model. We conclude the paper by
offering some discussions in section VI.

\sect{$U_q(gl(1|1))$ free fermion model}

\subsection{The background of the model }

Let $V$ be the 2-dimensional $U_q(gl(1|1))$-module and $R\in
End(V\otimes V)$ the $R$-matrix associated with this module. $V$
is $Z_2$-graded, and in the following we choose the FB grading for
$V$, i.e. $[1]=1,[2]=0$. The $R$-matrix depends on the difference
of two spectral parameters $u_1$ and $u_2$ associated with the two
copies of $V$, and is, in the FB grading, given by
\begin{eqnarray}
 R_{12}(u_1,u_2)=R_{12}(u_1-u_2)
          &=&\left(\begin{array}{ccccccccc}
 c_{12}&0&0 & 0\\
 0&a_{12}& b^+_{12}&0 \\
 0&b^-_{12}&a_{12}&0\\
 0&0&0&1
 \end{array} \right), \label{de:R} \nonumber\\
\end{eqnarray}
where
\begin{eqnarray}
&&
a_{12}=a(u_1,u_2)\equiv
  {\sinh(u_1-u_2)\over\sinh(u_1-u_2+\eta)},\quad \quad
b^\pm_{12}=b^\pm(u_1,u_2)\equiv
  {e^{\pm (u_1-u_2)}\sinh\eta\over\sinh(u_1-u_2+\eta)},\quad\quad
\nonumber\\
&&c_{12}=c(u_1,u_2)\equiv
{\sinh(u_1-u_2-\eta)\over\sinh(u_1-u_2+\eta)}
\end{eqnarray}
with $\eta\in C$ being the crossing parameter. One can easily
check that the $R$-matrix satisfies the unitary relation
\begin{equation}
R_{21}R_{12}=1.
\end{equation}
Here and throughout $R_{ij}\equiv R_{ij}(u_i,u_j)$. The $R$-matrix
satisfies the graded Yang-Baxter equation (GYBE)
\begin{equation}
R_{12}R_{13}R_{23}=R_{23}R_{13}R_{12}.
\end{equation}
In terms of the matrix elements defined by
\begin{equation}
R(u)(v^{i'}\otimes v^{j'})=\sum_{i,j}R(u)^{i'j'}_{ij}(v^{i}\otimes
v^{j}),
\end{equation}
the GYBE reads
\begin{eqnarray}
&&
\sum_{i',j',k'}R(u_1-u_2)^{i'j'}_{ij}R(u_1-u_3)^{i''k'}_{i'k}R(u_2-u_3)^{j''k''}_{j'k'}
    (-1)^{[j']([i']+[i''])}\nonumber\\
&=&\sum_{i',j',k'}R(u_2-u_3)^{j'k'}_{jk}R(u_1-u_3)^{i'k''}_{ik'}R(u_1-u_2)^{i''j''}_{i'j'}
    (-1)^{[j']([i]+[i'])}.
\end{eqnarray}

The quantum monodromy matrix $T(u)$ of the free fermion model on a
lattice of length $N$ is defined as
\begin{eqnarray}
T_0(u)=R_{0N}(u,z_N)R_{0N-1}(u,z_{N-1})_{\ldots} R_{01}(u,z_1),
\label{de:T}
\end{eqnarray}
where the index 0 refers to the auxiliary space  and $\{z_i\}$ are
arbitrary inhomogeneous parameters depending on site $i$. $T(u)$
can be represented in the auxiliary space as the $2\times 2$
matrix whose elements are operators acting on the quantum space
$V^{\otimes N}$:
\begin{eqnarray}
 T_0(u)=\left(\begin{array}{ccc}
 A(u)&B(u)\\C(u)&D(u)
 \end{array}\right)_{(0)}. \label{de:T-marix}
 \end{eqnarray}
 By using the GYBE, one may prove that the monodromy matrix
satisfies the GYBE
\begin{eqnarray}
R_{12}(u-v)T_1(u)T_2(v)=T_2(v)T_1(u)R_{12}(u-v).\label{eq:GYBE}
\end{eqnarray}
or in matrix form,
\begin{eqnarray}
&&\sum_{i',j'}R(u-v)^{i'j'}_{ij}
 T(u)^{i''}_{i'}T(v)^{j''}_{j'}(-1)^{[j']([i']+[i''])}
    \nonumber\\ && \mbox{} \quad\quad
=\sum_{i',j'}T(v)^{j'}_{j}T(u)^{i'}_{i}
R(u-v)^{i''j''}_{i'j'}(-1)^{[j']([i]+[i'])}.
\end{eqnarray}

Define the transfer matrix $t(u)$
\begin{eqnarray}
t(u)=str_0T_0(u),\label{de:t}
\end{eqnarray}
where $str_0$ denotes the supertrace over the auxiliary space.
With the help of the GYBE, one may check that the transfer matrix
satisfies the commutation relation $[t(u),t(v)]=0,$ ensuring the
integrability of the system. The transfer matrix gives the
Hamiltonian of the system:
\begin{eqnarray}
H&=&{d\ln t(u)\over du}|_{u=0}\nonumber\\
 &=&{1\over \sinh\eta} \sum_{j=1}^N\left(
     E^{12}_{(j)}E^{21}_{(j+1)}
   + E^{21}_{(j)}E^{12}_{(j+1)}
   - 2\cosh\eta E^{11}_{(j)}E^{11}_{(j+1)}\right.
   \nonumber\\
 &&\mbox{}\quad\left.
   - (e^{\eta}E^{11}_{(j)}E^{22}_{(j+1)}
     +e^{-\eta}E^{22}_{(j)}E^{11}_{(j+1)})\right),
     \label{de:H}
\end{eqnarray}
where $E_{(k)}^{ij}$ are generators, which act on the $k$th space,
of the superalgebra $U_q(gl(1|1))$.

Using the standard fermionic representation
\begin{eqnarray}
E_{(k)}^{12}=c_k,\quad E_{(k)}^{21}=c^\dag_k,\quad
E^{11}_{(k)}=1-n_k,\quad E^{22}_{(k)}=n_k, \quad n_k=c^\dag_kc_k,
\end{eqnarray}
the Hamiltonian can be rewritten as
\begin{eqnarray}
H={1\over \sinh\eta} \sum_{j=1}^N\left(
c_jc_{j+1}^\dag+c^\dag_jc_{j+1}-2\cosh\eta(1-n_j)\right).
\label{eq:H}
\end{eqnarray}

\subsection{Algebraic Bethe ansatz}

The transfer matrix (\ref{de:t}) can be diagonalized by using the
algebra Bethe ansatz. Define the Bethe state of the system
\begin{eqnarray}
\Phi_N(v_1,v_2,\ldots,v_n)=\prod_{i=1}^nC(v_i)|0>,
\label{de:Bethe-state}
\end{eqnarray}
where $|0>$ is the pseudo-vacuum,
\begin{eqnarray}
|0>=\prod_{k=1}^N
 \left(\begin{array}{c}0\\1\end{array}\right)_{(k)}
 \label{de:vacuum}
\end{eqnarray}
and the index $(k)$ indicates the $k$th space.

Applying the elements of the monodromy matrix (\ref{de:T-marix})
to the pseudo-vacuum $|0>$ and its dual, we easily obtain
\begin{eqnarray}
&&B(u)|0>=0,\quad <0|C(u)=0,\quad D|0>=|0>,\quad <0|D(u)=<0|,
\quad \nonumber\\
 && A(u)|0>=\prod_{i=1}^Na(u,z_i)|0>,\quad
 <0|A(u)=\prod_{i=1}^Na(u,z_i)<0|.
\end{eqnarray}

With the help of the GYBE (\ref{eq:GYBE}), we obtain the
commutation relations between the elements of the monodromy matrix
\begin{eqnarray}
C(u)C(v)&=&-c(u,v)C(v)C(u), \label{eq:commu-CC}\\
D(u)D(v)&=&D(v)D(u),\label{eq:commu-DD}\\
 A(u)C(v)&=&{c(u,v)\over a(u,v)}C(v)A(u)
 +{b^+(u,v)\over a(u,v)}C(u)A(v), \label{eq:commu-AC}\\
D(u)C(v)&=&{1\over a(v,u)}C(v)D(u)
 -{b^-(v,u)\over a(v,u)}C(u)D(v), \label{eq:commu-DC}\\
B(u)C(v)&=&-C(v)B(u)+{b^+(u,v)\over a(u,v)}
 [D(v)A(u)-D(u)A(v)]\nonumber\\
 &=&-C(v)B(u)+{b^+(u,v)\over a(u,v)}
 [D(u)t(v)-D(v)t(u)]. \label{eq:commu-BC}
\end{eqnarray}

Thus applying the transfer matrix $t(u)=D(u)-A(u)$ to the Bethe
state and using the commutation relations repeatedly, we obtain
the eigenvalues of $t(u)$ as
\begin{eqnarray}
t(u)\Phi_N=\Lambda(u,\{v_k\})\Phi_N=
 \left[\prod_{k=1}^n{1\over a(v_k,u)}
 -\prod_{j=1}^N a(u,z_j)\prod_{k=1}^n{c(u,v_k)\over
 a(u,v_k)}\right]\Phi_N   \label{eq:Lambda}
\end{eqnarray}
providing  $v_k\, (k=1,2,\ldots,n)$ satisfying the Bethe ansatz
equations (BAE)
\begin{eqnarray}
\prod_{j=1}^Na(v_k,z_j)=1. \label{eq:BAE}
\end{eqnarray}

For late use, we define the state of the free fermion chain of
length $N$
\begin{eqnarray}
|a_1a_2\ldots a_N>=|a_1>_{(1)}|a_2>_{(2)}\ldots |a_N>_{(N)}
\end{eqnarray}
and its dual
\begin{eqnarray}
|a_1a_2\ldots a_N>^\dag=<a_N|_{(N)}<a_{N-1}|_{(N-1)}\ldots
<a_1|_{(1)}
 \equiv <a_Na_{N-1}\ldots a_1|.
\end{eqnarray}


\sect{Drinfeld twists of the model}
\subsection{Factorizing $F$-matrix and its inverse}
Following \cite{Maillet96}, we now introduce the notation
$R^\sigma_{1\ldots N}$, where $\sigma$ is any element of the
permutation group ${\cal S}_N$. We note that we may rewrite the
GYBE as
\begin{eqnarray}
R_{23}^{\sigma_{23}}T_{0,23}=T_{0,32}R_{23}^{\sigma_{23}},
    \label{eq:RT-sigma23}
\end{eqnarray}
where $T_{0,23}\equiv R_{03}R_{02}$ and $\sigma_{23}$ is the
transposition of space labels (2,3). It follows that $R_{1\ldots
N}^{\sigma}$ is a product of elementary $R$-matrices
\cite{Maillet96,zsy04}, corresponding to a decomposition of
$\sigma$ into elementary transpositions. With the help of the
GYBE, one may generalize (\ref{eq:RT-sigma23}) to a $N$-fold
tensor product of spaces
\begin{eqnarray}
 R_{1\ldots N}^{\sigma}T_{0,1\ldots N}
  =T_{0,\sigma(1\ldots N)}R_{1\ldots N}^{\sigma},
     \label{eq:RT-sigma}
\end{eqnarray}
where $T_{0,1\ldots N}\equiv R_{0N}\ldots R_{01}.$ This implies
the ``decomposition" law
\begin{eqnarray}
 R^{\sigma'\sigma}_{1\ldots N}
 =R^{\sigma}_{\sigma'(1\ldots N)}
  R^{\sigma'}_{1\ldots N},\label{eq:R-RR}
\end{eqnarray}
for a product of two elements in ${\cal S}_N$. Note that
$R^{\sigma}_{\sigma'(1\ldots N)}$ satisfies the relation
\begin{eqnarray}
 R^{\sigma}_{\sigma'(1\ldots N)}
 T_{0,\sigma'(1\ldots N)}
 =T_{0,\sigma'\sigma(1\ldots N)}
  R^{\sigma}_{\sigma'(1\ldots N)}. \label{eq:RT-sigma'}
\end{eqnarray}
Let us write the elements of $R_{1\ldots N}^{\sigma}$ as
\begin{eqnarray}
 \left(R_{1\ldots N}^{\sigma}\right)
  ^{\alpha_{\sigma(N)}\ldots \alpha_{\sigma(1)}}
  _{\beta_N\ldots \beta_1},
\end{eqnarray}
where the labels in the upper indices are permuted relative to the
lower indices according to $\sigma$.

We proved in \cite{zsy04,zsy0502,zsy0503} that for the $R$-matrix
$R^\sigma_{1\ldots N}$, there exists a non-degenerate
lower-diagonal $F$-matrix (the Drinfeld twist) satisfying the
relation
\begin{eqnarray}
F_{\sigma(1\ldots N)}(z_{\sigma(1)},\ldots,z_{\sigma(N)})
  R_{1\ldots N}^\sigma(z_1,\ldots,z_N)
 =F_{1\ldots N}(z_1,\ldots,z_N). \label{eq:R-F-N}
\end{eqnarray}
Explicitly,
\begin{eqnarray}
F_{1,\ldots N}=\sum_{\sigma\in {\cal S}_N}
   \sum_{\alpha_{\sigma(1)}\ldots\alpha_{\sigma(N)}}^{\quad\quad *}
   \prod_{j=1}^N P_{\sigma(j)}^{\alpha_{\sigma(j)}}
   S(c,\sigma,\alpha_\sigma)R_{1\ldots N}^\sigma, \label{de:F}
\end{eqnarray}
where the sum $\sum^*$  is over all non-decreasing sequences of
the labels $\alpha_{\sigma(i)}$:
\begin{eqnarray}
&& \alpha_{\sigma(i+1)}\geq \alpha_{\sigma(i)},
 \quad\mbox{if}\quad
              \sigma(i+1)>\sigma(i), \nonumber\\
&& \alpha_{\sigma(i+1)}> \alpha_{\sigma(i)},
 \quad \mbox{if}\quad
              \sigma(i+1)<\sigma(i) \label{cond:F}
\end{eqnarray}
and the c-number function $S(c,\sigma,\alpha_\sigma)$ is given by
\begin{eqnarray}
S(c,\sigma,\alpha_\sigma)\equiv \exp\left\{{1\over2}
 \sum_{l>k=1}^N\left(1-(-1)^{[\alpha_{\sigma(k)}]}\right)
 \delta_{\alpha_{\sigma(k)},\alpha_{\sigma(l)}}
    \ln(1+c_{\sigma(k)\sigma(l)})\right\}.
\end{eqnarray}
The inverse of the $F$-matrix is given by
\begin{equation}
F^{-1}_{1\ldots N}=F^*_{1\ldots N}\prod_{i<j}\Delta_{ij}^{-1}
\label{eq:F-inverse}
\end{equation}
with
\begin{eqnarray}
\Delta_{ij}=\mbox{diag}\left((1+c_{ij})(1+c_{ji}),a_{ji},a_{ij},1\right)
\end{eqnarray}
and
\begin{eqnarray}
F^*_{1\ldots N}&=&\sum_{\sigma\in {\cal S}_N}
   \sum_{\alpha_{\sigma(1)}\ldots\alpha_{\sigma(N)}}^{\quad\quad **}
   S(c,\sigma,\alpha_\sigma)R_{\sigma(1\ldots N)}^{\sigma^{-1}}
   \prod_{j=1}^N P_{\sigma(j)}^{\alpha_{\sigma(j)}},
    \label{de:F*}  \nonumber\\
\end{eqnarray}
where the sum $\sum^{**}$ is taken over all possible $\alpha_i$
which satisfies the following non-increasing constraints:
\begin{eqnarray}
&& \alpha_{\sigma(i+1)}\leq \alpha_{\sigma(i)},
 \quad\mbox{if}\quad
              \sigma(i+1)<\sigma(i), \nonumber\\
&& \alpha_{\sigma(i+1)}< \alpha_{\sigma(i)},\quad \mbox{if}\quad
              \sigma(i+1)>\sigma(i). \label{cond:F*}
\end{eqnarray}

\subsection{Symmetric representation of the Bethe state}
The non-degeneracy of the $F$-matrix means that its column vectors
form a complete basis, which is called the $F$-basis. By the
procedure in \cite{zsy0503}, we find that in the $F$-basis, the
simple generators of the superalgebra $U_q(gl(1|1))$ have the
symmetric form:
\begin{eqnarray}
\tilde E^{12}
 &=&F_{12\ldots N}E^{12}F^{-1}_{12\ldots N}\nonumber\\
 &=&\sum_{i=1}^NE^{12}_{(i)}\otimes_{j\ne i}\mbox{diag}
     \left(2e^{-\eta}\cosh \eta, e^{-\eta}\right)_{(j)},\\
\tilde E^{21}&=&F_{12\ldots N}E^{21}F^{-1}_{12\ldots N}\nonumber\\
 &=&\sum_{i=1}^NE^{12}_{(i)}\otimes_{j\ne i}\mbox{diag}
     \left(e^{\eta}(2a_{ji}\cosh \eta)^{-1},
      e^{\eta}a^{-1}_{ji}\right)_{(j)}.
\end{eqnarray}
Similarly, the diagonal element $D(u)$ of the monodromy matrix in
the $F$-basis is given by
\begin{eqnarray}
\tilde D(u)&=&F_{12\ldots N} D(u)F^{-1}_{12\ldots N}
 =\otimes_{j=1}^N\mbox{diag}\left(a_{oj},1\right),
\end{eqnarray}
where $a_{0j}\equiv a(u,z_j)$.

Then, the creation and annihilation operators $C(u)$ and $B(u)$
read, in the $F$-basis,
\begin{eqnarray}
\tilde C(u)&=&F_{12\ldots
N} C(u)F^{-1}_{12\ldots N}
 =(q^{-1}\tilde E^{12}_{(i)} \tilde D(u)
  -\tilde D(u)\tilde E^{12}_{(i)})q^{-\sum_{i=1}^Nh_{(i)}}
  \label{eq:C-tilde}
  \nonumber\\
 &=&\sum_{i=1}^Nb^-_{0i}E^{12}_{(i)}\otimes_{j\ne i}
 \mbox{diag}\left(2a_{0j}\cosh\eta,1\right)_{(j)},\\
 \tilde B(u)&=&F_{12\ldots N} B(u)F^{-1}_{12\ldots N}
=q^{\sum_{i=1}^N h_{(i)}}
  (\tilde E^{21}\tilde D-q\tilde D\tilde E^{21})\nonumber\\
 &=&-\sum_{i=1}^N b^+_{0i}E^{21}_{(i)}\otimes_{j\ne i}
  \mbox{diag}\left(a_{0j}(2a_{ji}\cosh\eta)^{-1},
     a_{ji}^{-1}\right)_{(j)}, \label{eq:B-tilde}
\end{eqnarray}
where $b^\pm_{0j}\equiv b^\pm(u,z_j)$, $q=e^\eta$ and $h\equiv
-E^{11} -E^{22}$.

Acting the $F$-matrix (\ref{de:F}) on the state (\ref{de:vacuum}),
one sees that the pseudo-vacuum is invariant. Therefore in the
$F$-basis, the Bethe state (\ref{de:Bethe-state}) becomes,
\begin{eqnarray}
\tilde \Phi_N(v_1,v_2,\ldots,v_n)
 \equiv F_{1\ldots N}\Phi_N(v_1,\ldots,v_n)
 =\prod_{i=1}^n \tilde C(v_n)|0>.\label{eq:phi-tilde}
\end{eqnarray}
Substituting (\ref{eq:C-tilde}) into (\ref{eq:phi-tilde}), we
obtain
\begin{eqnarray}
\tilde \Phi_N(v_1,\ldots,v_n)
 =(2\cosh\eta)^{{n(n-1)\over 2}}\sum_{i_1<\ldots< i_{n}}
B^-_{n}(v_1,\ldots,v_{n}|z_{i_1},\ldots,z_{i_{n}})
E_{(i_1)}^{12}\ldots E_{(i_{n})}^{12}\,|0>\;, \nonumber\\
 \label{Phi_2}
\end{eqnarray}
where
\begin{eqnarray}
B^\pm_{n}(v_1,\ldots,v_{n}|z_{i_1},\ldots,z_{i_{n}})
 &=&
 \sum_{\sigma\in {\cal S}_n}\mbox{sign}(\sigma)
  \prod_{k=1}^n b^\pm(v_k,z_{i_{\sigma(k)}})
  \prod_{l=k+1}^na(v_k,z_{i_{\sigma(l)}})\nonumber\\
  &=&\mbox{det}{\cal B}^\pm(\{v_k\},\{z_j\})
  \label{eq:B_n}
\end{eqnarray}
with ${\cal B}^\pm(\{v_i\},\{z_j\})$ being a $n\times n$ matrix
with matrix elements
\begin{equation}
{\cal B}_{\alpha\beta}^\pm
 =b^\pm(v_\alpha,z_\beta)
  \prod_{\gamma=1}^{\alpha-1}a(v_\gamma,z_\beta).
\end{equation}

Similarly, acting $\tilde B(u_n)\ldots\tilde B(u_1)$ on the dual
pseudo-vacuum state, we have,
\begin{eqnarray}
<0|\tilde B(u_n)\ldots\tilde B(u_1)
&=&(-1)^n(2\cosh\eta)^{{-n(n-1)\over 2}}\sum_{i_1<\ldots< i_{n}}
\prod_{l=1}^n\prod_{k=1,\ne i_l}^N
 a^{-1}(z_{k},z_{i_l})\nonumber\\
&&\times \mbox{det}{\cal B}^+(\{v_k\},\{z_{i_j}\})
<0|E_{(i_n)}^{21}\ldots E_{(i_{1})}^{21}\ \;.
\end{eqnarray}


\sect{Determinant representation of the scalar product of the
Bethe states}

In \cite{Korepin93,Kitanine98}, the authors gave the determinant
representation of the scalar product of the Bethe state for the
spin-1/2 XXZ model. In this section, we derive the determinant
representation of the scalar product of the $U_q(gl(1|1))$ Bethe
states defined by
\begin{eqnarray}
S_n(\{u_j\},\{v_k\})
 =<0|B(u_n)\ldots B(u_1)C(v_1)\ldots C(v_n)|0>. \label{de:S_n}
\end{eqnarray}
The $F$-invariance of the pseudo-vacuum state $|0>$ and its dual
state $<0|$ leads to
%
\begin{eqnarray}
S_n(\{u_j\},\{v_k\})
 =<0|\tilde B(u_n)\ldots\tilde  B(u_1)
  \tilde C(v_1)\ldots \tilde C(v_n)|0>. \label{de:S_n-tilde}
\end{eqnarray}

Following \cite{Kitanine98}, we define
\begin{eqnarray}
&&G^{(m)}(\{v_k\},u_1,\ldots,u_m,i_{m+1},\ldots,i_n)\nonumber\\
 &&=<i_{n},\ldots,i_{m+1}|\tilde B(u_m)\ldots \tilde B(u_1)
  \tilde C(v_1)\ldots \tilde C(v_n)|0>, \label{de:G-m}
\end{eqnarray}
where $i_k$ $(k=m+1,\ldots,n)$, ordered as $i_{m+1}<\ldots<i_n$,
indicate the positions having state
$\left(\begin{array}{c}1\\0\end{array}\right)$, and other
positions have state
$\left(\begin{array}{c}0\\1\end{array}\right)$ . One sees that
when $m=n$, $G^{(n)}=S_n$. Inserting a complete set and noticing
(\ref{eq:B-tilde}),
(\ref{de:G-m})
 becomes
\begin{eqnarray}
&&G^{(m)}(\{v_k\},u_1,\ldots,u_m,i_{m+1},\ldots,i_n)\nonumber\\
 &&=\sum_{j\ne i_{m+1},\ldots,i_n}^N
     <i_{n},\ldots,i_{m+1}|\tilde B(u_m)
     |i_{m+1},\ldots, i_{m+p},j,i_{m+p+1},\ldots,i_n> \nonumber\\
 &&\quad\times
 G^{(m-1)}(\{v_k\},u_1,\ldots,u_{m-1},i_{m+1},
 \ldots, i_{m+p},j,i_{m+p+1},\ldots,i_n).
 \label{de:Gm-Gm-1}
\end{eqnarray}
In view of (\ref{eq:B-tilde}), we have
\begin{eqnarray}
&&<i_{n},\ldots,i_{m+1}|\tilde B(u_m)
     |i_{m+1},\ldots, i_{m+p},j,i_{m+p+1},\ldots,i_n>\nonumber\\
 &&=-(2\cosh\eta)^{-(n-m)}\cdot(-1)^{p}\,
 b^+(u_m,z_j)\prod_{k\ne j}^Na^{-1}(z_k,z_j)
 \prod_{l=m+1}^n a(u_m,z_{i_l}). \label{eq:B-expect}
\end{eqnarray}

With the help of (\ref{Phi_2}), we obtain $G^{(0)}$:
\begin{eqnarray}
&& G^{(0)}(\{v_k\},i_1,\ldots,i_n)
 =<i_n,\ldots,i_1|\prod_{k=1}^n
 \tilde C(v_k)|0> \nonumber\\
 &=&(2\cosh\eta)^{{n(n-1)\over 2}}
 \mbox{det} {\cal B}^-(\{v_k\},\{z_{i_l}\}). \label{eq:G-0}
\end{eqnarray}

We now compute $G^{(1)}$ by using the recursion relation
(\ref{de:Gm-Gm-1}). Substituting (\ref{eq:B-expect}) and
(\ref{eq:G-0}) into (\ref{de:Gm-Gm-1}), we obtain
\begin{eqnarray}
&& G^{(1)}(\{v_k\},u_1,i_2,\ldots,i_n)\nonumber\\
&&  =\sum^N_{j\ne i_2,\ldots,i_n}<i_{n},\ldots,i_{2}|\tilde B(u_1)
     |i_{2},\ldots, i_{p+1},j,i_{p+2},\ldots,i_n> \nonumber\\
&&\quad\times
G^{(0)}(\{v_k\},i_2,\ldots,i_{p+1},j,i_{p+2},\ldots,i_n)
   \nonumber\\&&
 =-(2\cosh\eta)^{{(n-1)(n-2)\over 2}}
 \sum^N_{j\ne i_2,\ldots,i_n}(-1)^{p}\,
 b^+(u_1,z_j)\prod_{k\ne j}^Na^{-1}(z_k,z_j)
 \prod_{l=2}^n a(u_1,z_{i_l})\nonumber\\ &&\quad\times
 \mbox{det}{\cal B}^-(\{v_k\},z_{i_2},\ldots,z_{i_{p+1}},
      z_j,z_{i_{p+2}},\ldots,z_{i_n}),
      \quad (k=1,\cdots,n). \label{eq:G-1-1}
 %
\end{eqnarray}
Let $v_k\, (k=1,\dots,n)$ label the row and $z_l\,
(l=i_2,\ldots,j,\ldots,i_n) $ label the column of the matrix
${\cal B}^-$. From (\ref{eq:G-0}), one sees that the column
indices in (\ref{eq:G-1-1}) satisfy the sequence
$i_2<\ldots<j<\ldots<i_n$. Therefore, moving the column $j$ in the
matrix ${\cal B}^-$ to the first column, we have
\begin{eqnarray}
&& G^{(1)}(\{v_k\},u_1,i_2,\ldots,i_n)\nonumber\\
&&=-(2\cosh\eta)^{{(n-1)(n-2)\over 2}}
 \sum^N_{j\ne i_1,\ldots,i_n}
 b^+(u_1,z_j)\prod_{k\ne j}^Na^{-1}(z_k,z_j)
 \prod_{l=2}^n a(u_1,z_{i_l})\nonumber\\ &&\quad\quad \times
 \mbox{det}{\cal B}(\{v_k\},z_j, z_{i_2},\ldots,z_{i_n})
 \nonumber\\ &&
 =-(2\cosh\eta)^{{(n-1)(n-2)\over 2}}\,
  \mbox{det}({\cal B}^-)^{(1)}(\{v_k\},u_1, z_{i_2},\ldots,z_{i_n}),
 \label{eq:G-1}
\end{eqnarray}
where the matrix $({\cal B}^-)^{(1)}(\{v_k\},u_1,
z_{i_2},\ldots,z_{i_n})$ is given by
\begin{eqnarray}
({\cal B}_{\alpha\beta}^-)^{(1)}&=&a(u_1,z_{i_\beta}){\cal
B}^-_{\alpha\beta} \quad\quad
 \mbox{for}\,\,\beta\geq 2,  \label{eq:B-1-0}\\
({\cal B}^-_{\alpha 1})^{(1)}&=& \sum^N_{j\ne i_2,\ldots,i_n}
 b^+(u_1,z_j)b^-(v_{\alpha},z_j)
 \prod_{\gamma=1}^{\alpha-1}a(v_\gamma,z_j)
 \prod_{k=1,\ne j}^N a^{-1}(z_{k},z_j). \label{eq:B-1-1}
\end{eqnarray}
Using the properties of determinant, one finds that if
$j=i_2,\ldots,i_n$, the corresponding terms in (\ref{eq:B-1-1})
contribute zero to the determinant. Thus, we may rewrite
(\ref{eq:B-1-1}) as
\begin{eqnarray}
({\cal B}_{\alpha 1}^-)^{(1)}&=&\sum_{j=1}^N
 {e^{u_1-v_\alpha}\sinh^2\eta\over
  \sinh(u_1-z_j+\eta)\sinh(v_\alpha-z_j+\eta)}
 \prod_{\gamma=1}^{\alpha-1}
 {\sinh(v_\gamma-z_j)\over\sinh(v_\gamma-z_j+\eta)}
 \nonumber\\ &&\times
 \prod_{k=1,\ne j}^N
 {\sinh(z_{k}-z_j+\eta)\over\sinh(z_{k}-z_j) }
 \nonumber\\
 &\equiv& e^{u_1}f(u_1).\label{eq:B-1-2}
\end{eqnarray}
Thanks to the Bethe ansatz equation (\ref{eq:BAE}), we may
construct the function
\begin{eqnarray}
{\cal M}_{\alpha\beta}^\pm
 &=&e^{\mp u_\beta}g(u_\beta)\nonumber\\
 &=&
 {e^{\pm(v_\alpha-u_\beta)}\sinh\eta\over\sinh(v_\alpha-u_\beta)}
 \prod_{\gamma=1}^{\alpha-1}
  {\sinh(v_\gamma-u_\beta-\eta)\over\sinh(v_\gamma-u_\beta)}
  \left\{1-\prod_{k=1}^N
  {\sinh(u_\beta-z_k)\over\sinh(u_\beta-z_k+\eta)}
  \right\}.
 \label{eq:M-ab}
\end{eqnarray}
Comparing $f(u_1)$ in (\ref{eq:B-1-2}) with $g(u_1)$ in
(\ref{eq:M-ab}), one finds that as functions of $u_1$, they have
the same residues at the simple pole $u_1=z_j-\eta$ mod($i\pi$),
and that when $u_1\rightarrow \infty$, they are bounded. Moreover,
one may prove that the residues of $g(u_1)$ at $u_1=v_\nu\,
(\nu=1,\ldots,\alpha)$ are zero because $v_\nu$ are solutions of
the Bethe ansatz equation (\ref{eq:BAE}). Therefore, we have
\begin{eqnarray}
({\cal B}_{\alpha 1}^-)^{(1)}={\cal M}^-_{\alpha1}
 ={b^-(v_\alpha,u_1)\over a(v_\alpha,u_1)}
 \prod_{\gamma=1}^{\alpha-1}a^{-1}(u_1,v_\gamma)\left(
 1-\prod_{k=1}^N a(u_1,z_k)\right). \label{eq:M-a1}
\end{eqnarray}

Then, by using the function $G^{(0)},G^{(1)}$ and the intermediate
function (\ref{de:Gm-Gm-1}) repeatedly, we obtain $G^{(m)}$ as
\begin{eqnarray}
&&G^{(m)}(\{v_k\},u_1,\ldots,u_m,i_{m+1},\ldots,i_n) \nonumber\\
 &&=(-1)^m(2\cosh\eta)^{{n(n-1)-m(2n-m-1)\over 2}}
 \prod_{1\leq j<k\leq m}a^{-1}(u_k,u_j)\nonumber\\
 &&\quad\quad\times
 \mbox{det}({\cal B}^-)^{(m)}(\{v_k\},u_1,\ldots,u_m,
  i_{m+1},\ldots,i_n)
 \label{eq:G-m}
\end{eqnarray}
with the matrix elements 
\begin{eqnarray}
({\cal B}_{\alpha\beta}^-)^{(m)}&=&\prod_{k=1}^m
 a(u_k,z_{i_\beta}){\cal B}_{\alpha\beta}^-,
 \quad\quad \mbox{for } \beta>m,\nonumber\\
({\cal B}_{\alpha\beta}^-)^{(m)}&=&{\cal M}^-_{\alpha\beta},
 \quad\quad\quad\quad\quad\quad\ \ \mbox{for } \beta\leq m.
\end{eqnarray}
(\ref{eq:G-m}) can be proved by induction. Firstly from
(\ref{eq:G-1}), (\ref{eq:B-1-0}) and (\ref{eq:M-a1}),
(\ref{eq:G-m}) is true for $m=1$. Assume (\ref{eq:G-m}) for
$G^{(m-1)}$. Let us show (\ref{eq:G-m}) for general $m$.
Substituting $G^{(m-1)}$ and (\ref{eq:B-expect}) into intermediate
function (\ref{de:Gm-Gm-1}), we have
\begin{eqnarray}
&&G^{(m)}(\{v_k\},u_1,\ldots,u_m,i_{m+1},\ldots,i_n)\nonumber\\
 &&=
 \sum_{j\ne i_{m+1},\ldots,i_n}^N
     <i_{n},\ldots,i_{m+1}|\tilde B(u_m)
     |i_{m+1},\ldots, i_{m+p},j,i_{m+p+1},\ldots,i_n> \nonumber\\
 &&\quad\quad\times
 G^{(m-1)}(\{v_k\},u_1,\ldots,u_{m-1},i_{m+1},
 \ldots, i_{m+p},j,i_{m+p+1},\ldots,i_n)\nonumber\\
 &&=-(2\cosh\eta)^{-(n-m)}
 \sum_{j\ne i_{m+1},\ldots,i_n}^N
  b^+(u_m,z_j)\prod_{k\ne j}^Na^{-1}(z_k,z_j)
 \prod_{l=m+1}^n a(u_m,z_{i_l})\nonumber\\
 &&\quad\quad\times
 G^{(m-1)}(\{v_k\},u_1,\ldots,u_{m-1},j,i_{m+1},
 \ldots, i_n)\nonumber\\
 &&=(-1)^m(2\cosh\eta)^{{n(n-1)-m(2n-m-1)\over 2}}
 \prod_{1\leq j<k\leq m-1}a^{-1}(u_k,u_j)\nonumber\\
 &&\quad\quad\times
 \mbox{det}{\cal
 B'}^{(m)}(\{v_k\},u_1\ldots,u_m,i_{m+1},\ldots,i_n),
\end{eqnarray}
where the matrix elements ${\cal B'}^{(m)}_{\alpha\beta}$ are
given by
\begin{eqnarray}
{\cal B'}^{(m)}_{\alpha\beta}&=&\prod_{k=1}^m
 a(u_k,z_{i_\beta}){\cal B}^-_{\alpha\beta}
 \quad\quad \mbox{for } \beta>m,\nonumber\\
{\cal B'}^{(m)}_{\alpha\beta}&=&{\cal M}^-_{\alpha\beta}
 \quad\quad\quad\quad\quad\quad\quad \mbox{for } \beta<m,
 \nonumber\\
{\cal B'}^{(m)}_{\alpha m}&=&\prod_{i=1}^{m-1}
 a(u_i,z_{j})
 \sum_{j\ne i_{m+1},\ldots,i_n}
 b^+(u_m,z_j)b^-(v_{\alpha},z_j)
 \prod_{\gamma=1}^{\alpha-1}a(v_\gamma,z_j)\nonumber\\
 &&\times
 \prod_{k=1,\ne j}^N a^{-1}(z_{k},z_j).
\end{eqnarray}
Thus, by the procedure leading to $({\cal
B}_{\alpha\beta}^-)^{(1)}$, we can prove
\begin{eqnarray}
{\cal B'}^{(m)}_{\alpha m}=\prod_{i=1}^{m-1}
 a^{-1}(u_m,u_i){\cal M}^-_{\alpha m}.
\end{eqnarray}
Then one sees that ${\cal B'}^{(m)}_{\alpha\beta}
 ={\cal B}^{(m)}_{\alpha\beta}$.
Therefore we have proved that (\ref{eq:G-m}) holds for all $m$.

When $m=n$, we obtain the scalar product $S_n(\{u_i\},\{v_j\})$,
\begin{eqnarray}
S_n(\{u_i\},\{v_j\}) =(-1)^n\prod_{k>l}^n
 a^{-1}(u_k,u_l)\mbox{det}{\cal M}^-(\{v_j\},\{u_{i}\}),
 \label{eq:Sn-Mab}
\end{eqnarray}
where the matrix elements of ${\cal M}^-$ are given by
\begin{eqnarray}
{\cal M}^\pm_{\alpha\beta}
 ={b^\pm(v_\alpha,u_\beta)\over a(v_\alpha,u_\beta)}
 \prod_{\gamma=1}^{\alpha-1}
 a^{-1}(u_\beta,v_\gamma)
 \left(1-\prod_{k=1}^N a(u_\beta,z_k)\right).
 \label{eq:S-ab}
\end{eqnarray}
By using the expression of the eigenvalues of the system
(\ref{eq:Lambda}), the scalar product (\ref{eq:Sn-Mab}) can be
rewritten as
\begin{eqnarray}
S_n(\{u_i\},\{v_j\})=(-1)^n\prod_{k>l}^n
 a^{-1}(u_k,u_l)\mbox{det}\hat{\cal M}^-(\{v_j\},\{u_i\})
 \label{eq:S-t}
\end{eqnarray}
with the matrix $\hat{\cal M}^\pm$ being
\begin{eqnarray}
\hat{\cal M}^\pm_{\alpha\beta}
 =e^{\pm(v_\alpha-u_\beta)}\sinh(u_\beta-v_\alpha)
 \prod_{\mu\ne\alpha}a(v_\mu,u_\beta)
 \prod_{\gamma=1}^{\alpha-1}a^{-1}(u_\alpha,v_\gamma)
 {\partial\Lambda(u_\beta,\{v_\alpha\})\over\partial v_{\alpha}}.
\end{eqnarray}
{\bf Remark:} In the derivation of (\ref{eq:Sn-Mab}), the
parameters $v_i$ in the state $\tilde C(v_1)\ldots \tilde
C(v_n)|0>$ are required to satisfy the BAE (\ref{eq:BAE}).
However, the parameters $u_j$ $(j=1,\ldots,n)$ in the dual state
$<0|\tilde B(u_n)\ldots\tilde B(u_1)$ do not need to satisfy the
BAE.

On the other hand, if we compute the scalar product by starting
from the dual state $<0|B(v_n)\ldots B(v_1)$, then by using the
same procedure, we have
\begin{eqnarray}
S_n(\{v_i\},\{u_j\})
 &=&<0|\tilde B(v_n)\ldots \tilde B(v_1)
 \tilde C(u_1)\ldots \tilde C(u_n)|0>\nonumber\\
 &=&(-1)^n\prod_{k>l}^n a^{-1}(u_k,u_l)\,
 \mbox{det}{\cal M}^+(\{v_i\},\{u_{j}\}). \label{eq:M+}
\end{eqnarray}
In the above equation, we have assumed that $\{v_i\}$ satisfy the
BAE.
\\[3mm]

Noticing the BAE (\ref{eq:BAE}), one sees that the scalar product
$S_n(\{u_i\},\{v_j\})=0$ if both parameter sets $\{u_i\}$ and
$\{v_j\}\, (\{v_j\}\ne\{u_i\}$ $i,j=1,\ldots, n)$ in
(\ref{eq:Sn-Mab}) and (\ref{eq:M+}) satisfy the BAE.

Let $u_\alpha\rightarrow v_\alpha$ $(\alpha=1,\ldots,n)$ in
(\ref{eq:Sn-Mab}), we obtain the Gaudin formula for the norm of
the $U_q(gl(1|1))$ Bethe state.
\begin{eqnarray}
{\cal S}_n&=&S_n(\{v_j\},\{v_k\})
 =<0|B(v_n)\ldots B(v_1)C(v_1)\ldots C(v_n)|0>\nonumber\\
 &=&(-1)^n\sinh^{n}\eta
 \prod_{k>j}^n{\sinh^2(v_k-v_j+\eta)\over\sinh^2(v_k-v_j)}
 \left[\prod_{\alpha=1}^n
 {1\over v_\alpha-u_\alpha}
 \left(1-\prod_{l=1}^N{\sinh(u_\alpha-z_l)
 \over\sinh(u_\alpha-z_l+\eta)}
 \right)\right]_{u_\alpha\rightarrow v_\alpha}
 \nonumber\\
 &=&(-1)^n\sinh^{n}\eta
 \prod_{k>j}^n{\sinh^2(v_k-v_j+\eta)\over\sinh^2(v_k-v_j)}
 \left[\prod_{\alpha=1}^n{\partial\over\partial u_\alpha}
 \ln\left(\prod_{l=1}^N{\sinh(u_\alpha-z_l)
 \over\sinh(u_\alpha-z_l+\eta)}
 \right)\right]_{u_\alpha\rightarrow v_\alpha}
 \nonumber\\
 &=&(-1)^n\sinh^{2n}\eta
 \prod_{k>j}^n{\sinh^2(v_k-v_j+\eta)\over\sinh^2(v_k-v_j)}
 \prod_{\alpha=1}^n\sum_{l=1}^N
 {1\over\sinh(v_\alpha-z_l)\sinh(v_\alpha-z_l+\eta)},
\end{eqnarray}
where we have used the BAE (\ref{eq:BAE}).


\sect{Correlation functions}

Having obtained the scalar product and the norm, we are now in the
position to compute the k-point correlation functions of the
model. In general, a k-point correlation function is defined by
\begin{eqnarray}
F^{\epsilon^1,\ldots,\epsilon^k}_n
 =<0| B(u_n)\ldots
 B(u_1)\epsilon^1_{i_1}\ldots\epsilon^k_{i_k}
  C(v_1)\ldots C(v_n)|0>,
\end{eqnarray}
where $\epsilon^j_{i_j}$ stand for the local fermion operators
$c_{i_j},c^{\dag}_{i_j}$ or $n_{i_j}$, and the lower indices $i_j$
indicate the positions of the fermion operators.

The authors in \cite{Korepin99} proved that the local spin and
field operators of the fundamental graded models can be
represented in terms of monodromy matrix. Specializing to the
current system, we obtain
\begin{eqnarray}
c_j^\dag&=&\prod_{k=1}^{j-1}(-A(z_k)+D(z_k))\cdot
  B(z_j)\cdot\prod_{k=j+1}^{N}(-A(z_k)+D(z_k)),
   \label{eq:c-dag} \\
c_j&=&\prod_{k=1}^{j-1}(-A(z_k)+D(z_k))\cdot
  C(z_j)\cdot\prod_{k=j+1}^{N}(-A(z_k)+D(z_k)),\label{eq:c}\\
n_j&=&\prod_{k=1}^{j-1}(-A(z_k)+D(z_k))\cdot
  D(z_j)\cdot\prod_{k=j+1}^{N}(-A(z_k)+D(z_k)). \label{eq:n}
\end{eqnarray}

\subsection{One point functions}
In this subsection, we compute the one point functions for the
local operators $c_m^\dag,\,c_m$ and $n_m$, respectively.

We first calculate $c_m^\dag$ . Noticing that the Bethe state and
its dual are eigenstates of the transfer matrix under the
constraint of the BAE, we have, from (\ref{eq:c-dag}),
\begin{eqnarray}
&&F^-_n(\{u_j\},z_m,\{v_k\})\nonumber\\
 &=&<0|B(u_n)\ldots B(u_1)c_m^\dag
   C(v_1)\ldots C(v_{n+1})|0>\nonumber\\
 &=&\phi_{m-1}(\{u_j\})\phi^{-1}_m(\{v_k\})
  <0|B(u_n)\ldots  B(u_1)B(z_m)
   C(v_1)\ldots C(v_{n+1})|0>\nonumber\\
 &=&\phi_{m-1}(\{u_j\})\phi^{-1}_m(\{v_k\})
 <0|\tilde B(u_n)\ldots \tilde B(u_1)\tilde B(z_m)
 \tilde C(v_1)\ldots \tilde C(v_{n+1})|0>\nonumber\\
 &=&\phi_{m-1}(\{u_j\})\phi^{-1}_m(\{v_k\})
 S_{n+1}(u_n,\ldots,u_1,z_m,\{v_j\})\nonumber\\
 &=&(-1)^{n+1}\phi_{m-1}(\{u_j\})\phi^{-1}_m(\{v_k\})
 \prod_{k>j}^n a^{-1}(u_k,u_j)
 \prod_{l=1}^n a^{-1}(u_l,z_m)
  \nonumber\\&& \times
 \mbox{det}{\cal
 M}^-(\{v_j\},z_m,u_1,\ldots,v_n),\label{eq:F-}
\end{eqnarray}
where $\phi_i(\{u_j\})=\prod_{k=1}^i\prod_{l=1}^n a(u_l,u_k)$. As
mentioned in the remark of the previous section, $F^-_n=0$ if the
parameter set $\{u_i\}$ $(i=1,\ldots,n)$ is not a subset of
$\{v_j\}$ $(j=1,\ldots,n+1)$. When $\{u_i\}\subset\{v_j\}$,
(\ref{eq:F-}) can be simplified to a simple function. For example,
if $u_i=v_{i+1}$ $(i=1,\ldots,n)$, the one point function $F^-$
becomes
\begin{eqnarray}
&&F^-_n(v_{n+1},\ldots,v_2,z_m,v_1,\ldots,v_{n+1})\nonumber\\
&=&(-1)^{n+1}{\phi_{m-1}(\{u_j\})\over\phi_m(\{v_k\})}
{e^{-(v_1-z_m)}\sinh^{2n+1}\eta\over\sinh(v_1-z_m)}
\prod_{k>j=2}^{n+1}{\sinh^2(v_k-v_j+\eta)\over\sinh^2(v_k-v_j)}
\prod_{j=2}^{n+1}{\sinh(v_j-z_m+\eta)\over\sinh(v_j-z_m)}
\nonumber\\ &&
\times\prod_{j=2}^{n+1}{\sinh(v_j-v_1+\eta)\over\sinh(v_j-v_1)}
 \prod_{\alpha=2}^{n+1}\sum_{l=1}^N
 {1\over\sinh(v_\alpha-z_l)\sinh(v_\alpha-z_l+\eta)}.
\end{eqnarray}

Similarly, when $\{u_i\}\subset\{v_j\}$, we obtain the one-point
function involving the operator $c_m$:
\begin{eqnarray}
&&F^+_n(\{v_k\},z_m,\{u_j\})\nonumber\\
 &=&<0|B(v_{n+1})\ldots B(v_1)c_m
   C(u_1)\ldots C(u_{n})|0>\nonumber\\
 &=&\phi_{m-1}(\{v_j\})\phi^{-1}_m(\{u_k\})
 S_{n+1}(\{v_j\},z_m,u_1,\ldots,u_n)\nonumber\\
 &=&(-1)^{n+1}\phi_{m-1}(\{v_j\})\phi^{-1}_m(\{u_k\})
 \prod_{k>j}^n a^{-1}(u_k,u_j)
  \prod_{l=1}^n a^{-1}(u_l,z_m)
  \nonumber\\&& \times
  \mbox{det}{\cal
 M}^+(\{v_j\},z_m,u_1,\ldots,v_n).\label{eq:F+}
\end{eqnarray}
$F^+_n$ is non-vanishing if $\{u_i\}\subset\{v_j\}$. When
$\{u_i\}\subset\{v_j\}$, (\ref{eq:F+}) can also be simplified to a
simple function. In the case $u_i=v_{i+1}$ $(i=1,\ldots,n)$, the
one point function $F^+$ becomes
\begin{eqnarray}
&&F^+_n(v_{n+1},\ldots,v_2,z_m,v_1,\ldots,v_{n+1})\nonumber\\
&=&(-1)^{n+1}{\phi_{m-1}(\{v_j\})\over\phi_m(\{u_k\})}
{e^{(v_1-z_m)}\sinh^{2n+1}\eta\over\sinh(v_1-z_m)}
\prod_{k>j=2}^{n+1}{\sinh^2(v_k-v_j+\eta)\over\sinh^2(v_k-v_j)}
\prod_{j=2}^{n+1}{\sinh(v_j-z_m+\eta)\over\sinh(v_j-z_m)}
\nonumber\\ &&
\times\prod_{j=2}^{n+1}{\sinh(v_j-v_1+\eta)\over\sinh(v_j-v_1)}
 \prod_{\alpha=2}^{n+1}\sum_{l=1}^N
 {1\over\sinh(v_\alpha-z_l)\sinh(v_\alpha-z_l+\eta)}.
\end{eqnarray}

The one-point function involving the operator $n_m$ is defined by
\begin{eqnarray}
F^{n_m}_n(\{u_j\},z_m,\{v_k\})=<0|B(u_n)\ldots B(u_1)n_m
  C(v_1)\ldots C(v_{n+1})|0>.
\end{eqnarray}
Substituting (\ref{eq:n}) into the above equation and considering
the BAE, we have
\begin{eqnarray}
&&F^{n_m}_n(\{u_j\},z_m,\{v_k\})
 =<0| B(u_n)\ldots B(u_1)n_m
   C(v_1)\ldots C(v_n)|0>\nonumber\\
&=&{\phi_{m-1}(\{u_j\})\over \phi_{m-1}(\{v_k\})} <0|\tilde
  B(u_n)\ldots\tilde B(u_1)\tilde D(z_m)
  \tilde C(v_1)\ldots\tilde C(v_n)|0>.\label{eq:Fn}
\end{eqnarray}
With the help of (\ref{eq:commu-DC}), we see
\begin{eqnarray}
&& D(z_m)C(v_1)\ldots C(v_n)|0>\nonumber\\
&=&\prod_{k=1}^n a^{-1}(v_k,z_m) C(v_1)\ldots C(v_n)|0>\nonumber\\
&&\mbox{}
 -\sum_{j=1}^n{b^-(v_j,z_m)\over
a(v_j,z_m)}
 \prod_{l=1}^{j-1}{c(v_l,v_j)\over c(v_l,z_m)}
 \prod_{k=1,\ne j}^n a^{-1}(v_k,v_j)\nonumber\\ &&\quad\times
  C(v_1)\ldots C(v_{j-1}) C(z_m)
  C(v_{j+1})\ldots
  C(v_n)|0>.\nonumber\\ \label{eq:commu-DCn}
\end{eqnarray}
Therefore, substituting (\ref{eq:commu-DCn}) into (\ref{eq:Fn}),
we obtain
\begin{eqnarray}
&&F^{n_m}_n(\{u_j\},z_m,\{v_k\})\nonumber\\
 &=&{\phi_{m-1}(\{u_j\})\over\phi_{m-1}(\{v_k\})}
  \prod_{k=1}^n a^{-1}(v_k,z_m)S_n(\{u_i\},\{v_j\}) \nonumber\\
&&\mbox{}  -\sum_{j=1}^n{b^-(v_j,z_m)\over a(v_j,z_m)}
 \prod_{l=1}^{j-1}{c(v_l,v_j)\over c(v_l,z_m)}
 \prod_{k=1,\ne j}^n a^{-1}(v_k,v_j)
  S_{n}(\{u_i\},v_1,\ldots,v_{j-1},z_m,v_{j+1},\ldots,v_n)
  \nonumber\\
 &=&(-1)^{n}{\phi_{m-1}(\{u_j\})\over\phi_{m-1}(\{v_k\})}
  \prod_{k=1}^n a^{-1}(v_k,z_m)
  \prod_{k>j} a^{-1}(v_k,v_j)\nonumber\\&& \times
  \mbox{det}\left[M^+(\{u_i\},\{v_j\})
  -{\cal N}(\{u_i\},\{v_j\},z_m)\right],
\end{eqnarray}
where ${\cal N}$ is a rank-one matrix with the following matrix
elements
\begin{eqnarray}
{\cal N}_{\alpha\beta}(\{u_i\},\{v_j\},z_m)
 &=&{e^{u_\alpha-v_\beta}\sinh^2\eta\over
  \sinh(u_\alpha-z_m)\sinh(v_\beta-z_m+\eta)}
  \prod_{i=1}^{\alpha-1} {\sinh(z_m-u_i+\eta)\over\sinh(z_m-u_i)}.
\end{eqnarray}
In the above derivation, we have used the following property of
determinant: If ${\cal A}$ is an arbitrary $n\times n$ matrix and
${\cal B}$ is a rank-one $n\times n$ matrix, then the determinant
of ${\cal A}+{\cal B}$ is given by
\begin{eqnarray}
\mbox{det}({\cal A}+{\cal B})=\mbox{det}{\cal A}
 +\sum_{i=1}^n\mbox{det}{\cal A}^{(i)},
\end{eqnarray}
where
\begin{eqnarray*}
&&{\cal A}^{(i)}_{\alpha\beta}={\cal A}_{\alpha\beta} \quad\quad
 \mbox{ for } \beta\ne i,\\
&&{\cal A}^{(i)}_{\alpha i}={\cal B}_{\alpha i}.
\end{eqnarray*}

\subsection{Correlation function of two adjacent operators}
In the subsection, we compute the correlation function of two
adjacent operators $c_m$ and $c_{m+1}$ defined by
\begin{eqnarray}
F^{-+}_n(\{u_i\},z_m,z_{m+1},\{v_j\})
 =<0|B(u_n)\ldots B(u_1)c_mc^\dag_{m+1}
  C(v_1)\ldots C(v_n)|0>\ .\label{de:F-+}
\end{eqnarray}
Substituting (\ref{eq:c}) and (\ref{eq:c-dag}) into the above
definition and considering the fact $\prod_{k=1}^Nt(z_k)=1$, we
have
\begin{eqnarray}
 &&F^{-+}_n(\{u_i\},\{v_j\},z_m,z_{m+1})\nonumber\\
 &&={\phi_{m-1}(\{u_i\})\over\phi_{m+1}(\{v_j\})}
 <0|\tilde B(u_n)\ldots\tilde B(u_1)\tilde {C}(z_m)
  \tilde { B}(z_{m+1})
  \tilde C(v_1)\ldots\tilde C(v_n)|0>\ .\label{eq:F-+}
\end{eqnarray}
By using the commutation relation (\ref{eq:commu-BC}), we obtain
\begin{eqnarray}
&&{B}(z_{m+1})
  C(v_1)\ldots C(v_n)|0>
=(-1)^n C(v_1)\ldots C(v_n)
   {B}(z_{m+1})|0>\nonumber\\
&&+\sum_{j=1}^n(-1)^{j+1}{b^+(z_{m+1},v_j)\over a(z_{m+1},v_j)}
 C(v_1)\ldots C(v_{j-1})
  D(z_{m+1}) t(v_j)  C(v_{j+1}) C(v_n)|0>
 \nonumber\\
&&+\sum_{j=1}^n(-1)^{j}{b^+(z_{m+1},v_j)\over a(z_{m+1},v_j)}
 C(v_1)\ldots C(v_{j-1})
  t(z_{m+1})D(v_j) C(v_{j+1}) C(v_n)|0>,
 \label{eq:BCn}
\end{eqnarray}
where $\tilde t(u)\equiv F_{1\ldots N} t(u) F^{-1}_{1\ldots N}$.
On the rhs of the above equation, one easily finds that the first
term is zero. Using the BAE, one may check that the second term
also equals to zero. Therefore, only the third term survives on
the rhs of the above equation and we have
\begin{eqnarray}
&& {B}(z_{m+1})  C(v_1)\ldots C(v_n)|0>\nonumber\\
&&=\sum_{j=1}^n(-1)^{j}\,{b^+(z_{m+1},v_j)\over a(z_{m+1},v_j)}
 \prod_{k=j+1}^n a^{-1}(v_k,z_{m+1})
 \prod_{l=j+1}^n a^{-1}(v_l,v_j)\nonumber\\ &&\quad\quad\times
  C(v_1)\ldots C(v_{j-1}) C(v_{j+1}) C(v_n)|0>
 \nonumber\\ && \quad
 +\sum_{j=1}^n(-1)^{j+1} \,{b^+(z_{m+1},v_j)\over a(z_{m+1},v_j)}
 \prod_{k=j+1}^n a^{-1}(v_k,z_{m+1})
 \nonumber\\ && \quad\quad\times
 \sum_{l=j+1}^n{b^-(v_l,v_j)\over a(v_l,v_j)}
  \prod_{m=j+1}^{l-1}{c(v_m,v_l)\over c(v_m,v_j)}
 \prod_{i=j+1,\ne l}^n a^{-1}(v_i,v_l)\nonumber\\ &&\quad\quad\times
  C(v_1)\ldots C(v_{j-1}) C(v_{j+1})\ldots
  C(v_{l-1}) C(v_j) C(v_{l+1})\ldots
  C(v_n)|0>\nonumber\\
 &&\equiv \sum_{j=1}^n
  M_j C(v_1)\ldots C(v_{j-1}) C(v_{j+1}) C(v_n)|0>
  \nonumber\\ &&\quad +\sum_{j=1}^n\sum_{l=j+1}^n M_{j,l}
  C(v_1)\ldots C(v_{j-1}) C(v_{j+1})\ldots
  C(v_{l-1}) C(v_j) C(v_{l+1})\ldots
  C(v_n)|0>.\nonumber\\ \label{eq:M-BCn}
\end{eqnarray}
Substituting (\ref{eq:M-BCn}) into (\ref{eq:F-+}), we obtain
two-point correlation function $F^{-+}_n$
\begin{eqnarray}
 &&F^{-+}_n(\{u_i\},z_m,z_{m+1},\{v_j\})\nonumber\\
 &&={\phi_{m-1}(\{u_i\})\over\phi_{m+1}(\{v_j\})}
 \left[\sum_{j=1}^n M_j S_n(\{u_i\},z_m,v_1,\ldots, v_{j-1},v_{j+1},v_n)
 \right. \nonumber\\ &&\left. \quad +
 \sum_{j=1}^n\sum_{l=j+1}^n M_{j,l}
 S_n(\{u_i\},z_m,v_1,\ldots, v_{j-1},v_{j+1},
 \ldots,v_{l-1},v_j, v_{l+1},\ldots,v_n)
 \right]
 \ .
\end{eqnarray}\\[5mm]


\sect{Discussion}

In this paper, with the help of the factorizing $F$-matrix
($F$-basis), we have obtained the determinant representations of
the scalar products and correlation functions of the
$U_q(gl(1|1))$ free fermion model.

In \cite{McCoy68}-\cite{Kitanine02}, the authors studied the
correlation functions of the free fermion model based on the
finite XX0 spin chain (XY model \cite{Lieb61}) with periodic
boundary condition
\begin{eqnarray}
H_{XX0}=\sum_{j=1}^N\left(\sigma_j^x\sigma_{j+1}^x
+\sigma_j^y\sigma_{j+1}^y +h\sigma^z_j\right), \label{eq:XX0}
\end{eqnarray}
where $\sigma^{\epsilon}\, (\epsilon=x,y,z)$ are the Pauli
matrices and $h$ is an external classical magnetic field. The
equivalence between the free fermion model and the XX0 model can
be proved by using the Jordan-Wigner transform
\begin{eqnarray}
c_k=\exp[i\pi Q_{k-1}]\sigma^+_k,\label{de:c-s}\\
c_k^{\dag}=\sigma^-_k\exp[i\pi Q_{k-1}],\label{de:c-s-dag}
\end{eqnarray}
where $\sigma^\pm={1\over 2}(\sigma^x\pm \sigma^y)$,
$Q_k=\sum_{j=1}^k{1\over 2}(1-\sigma^z_k)$. Because of the
periodic boundary condition of the finite XX0 spin chain, we have
\begin{equation}
\sigma^\pm_{N+1}=\sigma_1^\pm.
\end{equation}
Substituting the Jordan-Wigner transforms into the above relation,
we obtain
\begin{eqnarray}
 c_{N+1}=\exp[i\pi Q_{N}]c_1, \quad
 c^\dag_{N+1}=c_1^\dag\exp[i\pi Q_{N}].
\end{eqnarray}
Thus, comparing the above boundary condition with that of the
$U_q(gl(1|1))$ free fermion model (\ref{eq:H}), we find that the
free fermion model arising from the XX0 model has a twisted
boundary condition which depends on the operator
$\sigma^z=\sum_{i=1}^N\sigma^z_i$.

On the other hand, by means of the Jordan-Wigner transform, the
$U_q(gl(1|1))$ free fermion model is equivalent to a twisted XX0
model, and the one-point correlation functions (\ref{eq:F-}) and
(\ref{eq:F+}) give rise to the $m$-point correlation functions of
the twisted XX0 model. For example: substituting
(\ref{de:c-s-dag}) into (\ref{eq:F-}), we obtain
\begin{eqnarray}
&&F^-_n(\{u_j\},z_m,\{v_k\})\nonumber\\
 &=&<0| B(u_n)\ldots  B(u_1)c_m^\dag
   C(v_1)\ldots  C(v_{n+1})|0>\nonumber\\
 &=&<0| B(u_n)\ldots  B(u_1)
 \sigma^z_1\ldots\sigma^z_{m-1}\sigma^-_m
   C(v_1)\ldots  C(v_{n+1})|0>. \label{eq:F-XX0}
\end{eqnarray}


{\bf Acknowledgements:} This work was financially supported by the
Australian Research Council. S.Y. Zhao was supported by the UQ
Postdoctoral Research Fellowship.

\end{document}